\documentclass[twocolumn,prl]{revtex4-2}
\usepackage{graphicx}
\usepackage{amssymb}
\usepackage{epstopdf}
\usepackage{amsmath,bm}
\usepackage{bm}
\usepackage{multirow}

\usepackage{isomath}
\usepackage{footnote}
\usepackage[bottom]{footmisc}
\usepackage{color}
\makeatletter

\@ifundefined{textcolor}{}
{
 \definecolor{BLACK}{gray}{0}
 \definecolor{WHITE}{gray}{1}
 \definecolor{RED}{rgb}{1,0,0}
 \definecolor{GREEN}{rgb}{0,1,0}
 \definecolor{BLUE}{rgb}{0,0,1}
 \definecolor{CYAN}{cmyk}{1,0,0,0}
 \definecolor{MAGENTA}{cmyk}{0,1,0,0}
 \definecolor{YELLOW}{cmyk}{0,0,1,0}
}

\usepackage[colorlinks,
            linkcolor=black,
            anchorcolor=black,
            citecolor=black
            ]{hyperref}
\usepackage{ulem}           
            
\makeatother

\DeclareMathAlphabet\mathbfcal{OMS}{cmsy}{b}{n}

\begin{document}

\title{Topological Quantum Spin Hall Semimetals with Light}
\author{Karyn Le Hur}
\affiliation{CPHT, CNRS, École polytechnique, Institut Polytechnique de Paris, 91120 Palaiseau, France}

\begin{abstract}
We introduce a quantum spin Hall semimetal or Fermi liquid characterized with a $\mathbb{Z}_2$ topological invariant, measurable through circularly polarized light. We propose its engineering through two topological metallic band structures in crystals on the honeycomb lattice, with spin-orbit interaction, realizable through the interplay of a charge or spin density wave substrate and Zeeman effects, in between a quantum spin Hall and a quantum anomalous Hall insulator. These systems show topologically protected helical edge or photo-induced currents. 
\end{abstract}
\maketitle

{\color{blue} Introduction.---} Topological systems are at the heart of interest in physics e.g. through the quantum Hall and quantum anomalous Hall effects, topological insulators and topological superconductors \cite{QH,QiZhang,HasanKane,Bernevig}. This finds various applications in protected current flows at the edges and in quantum information through Majorana fermions \cite{ElliottFranz,Mi,MajoQuantumInfo}. These topological systems develop an energy gap at the Fermi energy in the bulk. The quest of topological semimetals is attracting a lot of attention in the community in three dimensions since the last decade with a relation to axion physics for Weyl semimetals \cite{SekineNomura}. This is stimulating important progress in sensing the quantum matter e.g. through circularly polarized light and the photogalvanic effect \cite{Moore}. Circularly polarized light is also an equivalent measure of the quantum Hall conductivity for two-dimensional (2D) quantum anomalous Hall insulators \cite{Goldman,Hamburg} and the light-induced information can be resolved from Dirac points on the honeycomb lattice when reaching the resonance \cite{Light1,Light2,Review}. The photogalvanic effect is similarly observed in topological insulators \cite{Herrero} and the information can yet be resolved at specific points within the Brillouin zone \cite{Light2,Review}. The quest for a two-dimensional topological semimetal phase is yet timely since the work of Young and Kane in 2015 \cite{YoungKane}, for a complete classification table. We have recently generalized the quantum anomalous Hall effect or Haldane model \cite{Haldane} to a protected 2D nodal-ring topological semimetal phase with applications in bilayer systems \cite{HH,KLH} and graphene \cite{LeHurSaati1,LeHurSaati2}, with related progress in the community \cite{BoFu}. Here, we introduce a quantum spin Hall semimetal in two dimensions with a topological Fermi surface, between a quantum spin Hall (QSH) insulator and a quantum anomalous Hall (QAH) insulator, when varying Zeeman effects leading to topologically protected $\mathbb{Z}_2$ Fermi liquids. Classifying topological metallic band structures, e.g. with the responses to light, is also timely since the proximity of Fermi surfaces generally leads to corrections in the quantum Hall responses \cite{Haldane2004,AlexKarynAndrew}. 

Through a microscopic model which can be realized, the band structures will show two topological  $\mathbb{Z}_2$ markers: one is quantized, referring to a $\mathbb{Z}_2$ invariant encoding some aspects of time-reversal symmetry and one is non-quantized encoding the spin magnetization of the ground state. Signatures of these two markers are measurable through circularly polarized light with a correspondence towards the Hall conductivities of different regions in the band structure. The total quantum Hall conductivity is zero. Then, we introduce an heterostructure with a magnetic substrate, instead of a charge density wave substrate, and we formulate an analogy between a QSH semimetal showing in this case a topological nodal ring and a model of two spheres \cite{HH} with a halved $\mathbb{Z}_2$ invariant corresponding to a pair of $\pi$ and $-\pi$ Berry phases (winding numbers). The recent article \cite{Dai} proposes a different scenario to realize a $\mathbb{Z}_2$ Weyl nodal-line topological semimetal related to the Kane-Mele model \cite{KaneMele}.

\begin{table*}[]
\begin{tabular}{||c|c|c||}
 \hline
 $K$ Dirac point & Eigenstates & Energetics  \\
 \hline
$\zeta=+1$ & $|-\rangle\otimes|-z\rangle$  & $-|{\bf d}_+|-r$  \\
$\zeta=+1$ & $|+\rangle\otimes|+z\rangle$ & $- |{\bf d}_{\mp}|+r$  \\
$\zeta=+1$ & $|-\rangle\otimes|+z\rangle$ & $|{\bf d}_{\mp}|+r$ \\
$\zeta=+1$ & $|+\rangle\otimes|-z\rangle$ & $|{\bf d}_+|-r$  \\
 \hline
    \end{tabular}
\begin{tabular}{||c|c|c||}
 \hline
  $K'$ Dirac point & Eigenstates & Energetics  \\
  \hline
$\zeta=-1$ & $|-\rangle\otimes|+z\rangle$  & $-|{\bf d}_{\pm}|+r$  \\
$\zeta=-1$ & $|+\rangle\otimes|-z\rangle$ & $-|{\bf d}_-|-r$  \\
$\zeta=-1$ & $|-\rangle\otimes|-z\rangle$ & $|{\bf d}_-|-r$  \\
$\zeta=-1$ & $|+\rangle\otimes|+z\rangle$ & $|{\bf d}_{\pm}|+r$  \\
 \hline
    \end{tabular}
    \caption{Eigenstates and corresponding energies at the two Dirac points for the two models. At these points, $d_x=d_y=0$, therefore $|{\bf d}_-|=d_z-M$ and $|{\bf d}_+|=d_z+M$.}
%\label{}
%\vskip -0.5cm
\end{table*}

{\color{blue} Topological Fermi Liquid with a QSH effect.---} The Hamiltonian on the honeycomb lattice takes the form 
\begin{equation}
H=\sum_{\bf k} \psi^{\dagger}({\bf k}) {\cal H}({\bf k})\psi({\bf k})
\end{equation}
with $\psi({\bf k})=(c_{A{\bf k}\uparrow},c_{B{\bf k}\uparrow},c_{A{\bf k}\downarrow},c_{B{\bf k}\downarrow})$ and
\begin{eqnarray}
\label{Hmodel}
{\cal H}({\bf k}) &=& d_z({\bf k})\sigma_z\otimes s_z + M\sigma_z\otimes \mathbb{I}+ d_{x}({\bf k}) \sigma_x\otimes \mathbb{I} \\ \nonumber
&+& d_{y}({\bf k})\sigma_y\otimes \mathbb{I} + r\mathbb{I}\otimes s_z.
\end{eqnarray}
The Hamiltonian is written with two sets of Pauli matrices: $\mathbfit{\sigma}$ acting on the Hilbert space $\{|+\rangle;|-\rangle\}$ associated to the two sublattices $A,B$, referring to the {\it pseudo-spin} space, and $\bf{s}$ acting on $\{|+\rangle_z;|-\rangle_z\}$ linked to the two {\it spin} polarizations of an electron. Here, the components $d_x$ and $d_y$ correspond to the kinetic term on the honeycomb lattice and $d_z({\bf k})$ encodes the presence of the Kane-Mele term such that ${\bf d}_z(-{\bf k})=-{\bf d}_z({\bf k})$ \cite{KaneMele,StephanKaryn}. Similar to the Haldane model on the honeycomb lattice \cite{Haldane,Light1} we can introduce $d_z=-d_z({\bf K})=d_z({\bf K}')=3\sqrt{3}t_2$ at the two Dirac points $K$ and $K'$, with $t_2$ corresponding then to the spin-orbit interaction \cite{StephanKaryn}, and write close to these points $d_z({\bf k})s_z\otimes\sigma_z=-\zeta d_z s_z\otimes \sigma_z$. Here, $\zeta=\pm 1$ at the $K$ and $K'$ Dirac points respectively. The quantum spin Hall effect is observed in a plethora of two-dimensional quantum materials \cite{WurzburgQSH,Reis,WTe2,WSe2,Bouchiat} and this specific form of spin-orbit interaction in the K,K' valleys can be realized in various ways; see e.g. Ref. \cite{RoserValenti}. A charge density wave substrate can produce the staggering potential term $M$ in the Hamiltonian, such that $M<d_z$ to favor the occurrence of topological phases, and a Zeeman effect is added along $z$ direction with $r>0$ such that $[{\cal H}({\bf k}),s_z]=0$. Doping topological materials with magnetic dopants is a way to produce such a Zeeman term with $r=\mu_B B_z$ with $B_z$ the magnetic field and $\mu_B$ the Bohr magneton; a transition from a quantum spin Hall to a quantum anomalous Hall insulator is e.g. observed in 2D Mercury \cite{Mercury} and Bismuth materials \cite{Bismuth} in this way. Another approach to engineer the $r$ term would be through a ferromagnetic material and an additional Hund coupling \cite{Guguchia,JulianKaryn}. It is also relevant to mention recent efforts to characterize
Bismuth thin films semimetals \cite{Tatara}.

For any wave-vector ${\bf k}$, the model can be classified in terms of the two eigenstates $|\psi_-\rangle$ and $|\psi_+\rangle$ in the pseudo-spin sector associated to a radial magnetic field structure
on the equivalent sphere model such that the north pole is equivalent to to the $K$ Dirac point and the south pole to $K'$, with specifically at the Dirac points $|\psi_-(K)\rangle=|-\rangle$, $|\psi_+(K)\rangle=|+\rangle$ \cite{Review}. Due to the form of the $d_z({\bf k})$  term we also have $|\psi_+(K)\rangle=|\psi_-(K')\rangle$ and $|\psi_-(K)\rangle=|\psi_+(K')\rangle$. The eigenstates in the spin sector are classified through the two spin eigenstates $|+z\rangle$ and $|-z\rangle$ associated to $s_z=\pm 1$. It is useful to introduce eigenstates at the two Dirac points where $d_x=d_y=0$ since the global topology can be generally encoded at these two specific points within the Brillouin zone \cite{Review}. A simple analysis of energetics leads to the eigenstates in Table 1.
in Table 1 with the correspondence $|\psi_-(K)\rangle=|-\rangle$ and $|\psi_+(K)\rangle=|+\rangle$. The possible energies are $\mp |{\bf d}_{\pm}|\mp r$, where ${\bf d}_{+}=(d_x,d_y,d_z+M)$ and 
${\bf d}_{-}=(d_x,d_y,d_z-M)$, respectively. In the $K$ valley, $\zeta=+$, the lowest-energy state is spin-polarized along $|-z\rangle$, corresponding to $|\psi_-\rangle\otimes|-z\rangle$ with energy $-|{\bf d}_+|-r$. The second energy band is related through the transformation $s_z\rightarrow -s_z$ and $\sigma_z\rightarrow -\sigma_z$, characterizing the QSH effect, and the eigenstate is $|\psi_+\rangle\otimes|+z\rangle$ associated to the energy $-|{\bf d}_-|+r$. When $r=0$ and $M<d_z$, the system is a QSH insulator which is in general robust towards interactions until the Mott transition \cite{StephanKaryn,HKH,WeiKarynStephan,Wurzburg,Sherbrooke}. When $r>d_z-M$, the second-energy band in the $K$ valley becomes metallic when the chemical potential resides at $\mu=0$, i.e. at half-filling. This refers to the metallic hole region in {\it dashed blue} on the (left) Figure. Through the modification $|{\bf d_-}|\leftrightarrow |{\bf d_+}|$ in the eigen-energies when navigating from one valley to the other, this energy band is now insulating at $K'$. The third-energy band in the valley close to the $K$ Dirac point is related to the second-energy band when flipping the pseudo-spin quantum number, with eigenstate $|\psi_-\rangle\otimes|+z\rangle$ and energy $|{\bf d}_-|+r$. On the Figure, $M\sim 2r<d_z$. The fourth branch with energy $|{\bf d}_+|-r$ reveals the eigenstate $|\psi_+\rangle\otimes|-z\rangle$. At $K'$, if we modify $|{\bf d}_+|\rightarrow |{\bf d}_-|$ this branch now participates to the ground state through the {\it dashed red} electron pocket, i.e. Fermi liquid region with a quadratic spectrum, as long as $d_z-M<r<d_z+M$. A similar realm of applicability occurs for the 2D quantum anomalous Hall semimetal \cite{HH,LeHurSaati1,LeHurSaati2}.

These two metallic particle and hole regions crossing the Fermi energy satisfy the equation $r=|{\bf d}_-|=\sqrt{d_x^2+d_y^2 +(d_z-M)^2}$ with $d_x^2+d_y^2=\hbar^2|{\bf p}|^2$ where ${\bf p}$ corresponds to a wave-vector deviation from K,K', 
and through the action of time-reversal symmetry is equivalent to a nodal ring structure around the $K$ point. In the quantum spin Hall effect, time-reversal symmetry can be introduced through the operator $U=i(\mathbb{I}\otimes s_y)\theta$ with $U^2=-1$ such that $-i s_y|+z\rangle = |-z\rangle$, $-i s_y|-z\rangle = -|+z\rangle$ and $\theta|\psi({\bf k})\rangle = |\psi(-{\bf k})\rangle^*$. Flipping the spin eigenstate under time-reversal symmetry also implies to flip the pseudo-spin eigenstate. In the present situation, the action of time-reversal symmetry on the Fermi surface is as follows. If we apply time-reversal symmetry on the eigenstate $|\psi_+\rangle\otimes|-z\rangle$ with energy $|{\bf d}_-|-r$ at point $K'$ this corresponds to a teleported eigenstate $|\psi_+\rangle\otimes|+z\rangle$ at point $K$ with the same energy $|{\bf d}_-|-r$, corresponding to the {\it orange} Fermi arc (pocket). Through the action of time-reversal symmetry the topological properties of the Fermi surface are equivalent to a unique energy band $|\psi_+\rangle\otimes|+z\rangle$ spreading on the whole Brillouin zone. This justifies why the Fermi surface is topological in the present case and characterized through a $\mathbb{Z}$ invariant due to the fact that $|\psi_+(K')\rangle = |\psi_-(K)\rangle$ forming a Skyrmion on the analogous sphere representation \cite{Review}. Since the two energy bands $|\psi_+\rangle\otimes|+z\rangle$ and $|\psi_-\rangle\otimes|-z\rangle$ are related through $s_z\rightarrow -s_z$ and $\sigma_z\rightarrow -\sigma_z$ we can then introduce a $\mathbb{Z}_2$ topological invariant and the term {\it `topological quantum spin Hall' semimetal}, that we develop hereafter through the quantum Hall transport and circularly polarized light; see Supplementary Material \cite{SM} for additional information. We will also address its robustness towards interactions and disorder.

\begin{figure*}[]\centering
\includegraphics[width=1\textwidth]{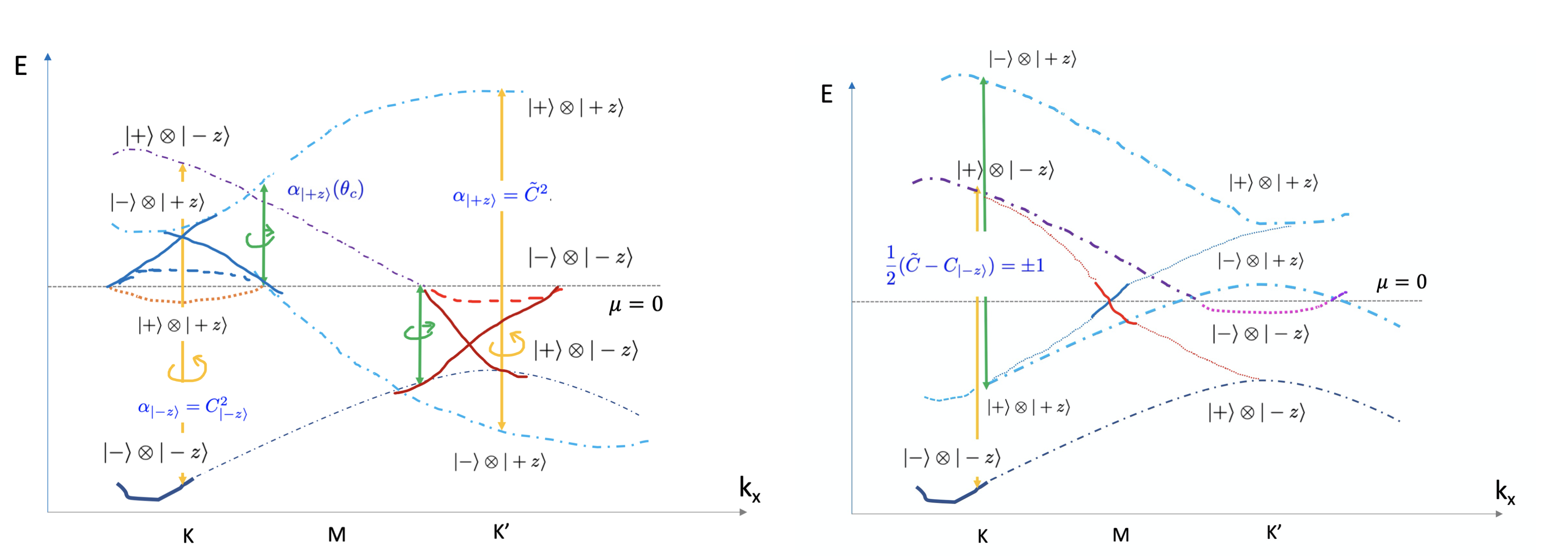} 
\vskip -0.3cm
\caption{Band Structures of the Quantum Spin Hall Semimetals. On the left, the response to the yellow circularly polarized light $(-)$ at $K$ and $K'$ reveals the ${\mathbb{Z}}_2$ invariant structure associated to Eq. (\ref{formula1}). This invariant can also be observed through transport when adjusting gently the chemical potential on both sides of the line at $\mu=0$. Shining circularly polarized light with the right green wave $(+)$ at the point(s) along the path crossing the line with zero energy reveals the sum of the square of the quantum Hall conductivities of the partially filled bands (domains). On the right, the quantum spin Hall semimetal is equivalently characterized through the halved $\mathbb{Z}_2$ invariant $\pm 1$ which can be resolved locally through light at the $K$ Dirac point.}
\label{Figuretopo}
\end{figure*}

{\color{blue} $\mathbb{Z}_2$ topological markers and transport.---}  For the fully occupied band $|\psi_-\rangle\otimes|-z\rangle$, the quantum Hall response agrees with the topological Chern number $C_{|-z\rangle}=-1$. 
This gives rise to a quantum Hall response 
\begin{equation}
\sigma_{xy}^{|-z\rangle,(1)} = - \frac{e^2}{h} = \frac{e^2}{h}C_{|-z\rangle}.
\end{equation}
When integrating the Berry curvatures on the two partially filled bands, then we reveal the quantized quantum Hall response for the Fermi surface 
\begin{equation}
\sigma_{xy}^{|+z\rangle}+ \sigma_{xy}^{|-z\rangle,(2)} = \frac{e^2}{h}\tilde{C}=-\frac{e^2}{h}C_{|-z\rangle},
\label{equationtransport}
\end{equation}
with $\tilde{C}=+1$. For the sake of clarity, we present the evaluation of this result in the Supplementary Material \cite{SM} from a mapping onto the sphere which allows for analytical elegant proofs including the responses to light. The quantum Hall response $\sigma_{xy}^{|+z\rangle}$ is associated to the filled (occupied
dashed-dotted light blue) region at zero temperature with spin polarization $+z$ and  $\sigma_{xy}^{|-z\rangle,(2)}$ corresponds to the quantum Hall response of the red dashed occupied region with fermions polarized along $-z$ around the $K'$ Dirac point. 
When mapping the 2D Brillouin zone onto a sphere, 
${\bf k}\rightarrow (\varphi_{\bf k},\tilde{\theta}_{\bf k})$, with $\varphi$ the azimuthal angle and $\tilde{\theta}$ the polar angle, the superscript symbol `tilde' in $\tilde{\theta}$ means that we absorb the effect of the $M$ term in a definition of the polar angle \cite{SM}. When different regions of the Fermi surface are characterized through orthogonal spin quantum numbers, we can integrate the Berry curvature on each region associated to a spin polarization and obtain the quantum Hall responses \cite{KLH,LeHurSaati2}. In the present situation, the band $|\psi_+\rangle\otimes|+z\rangle$ participates to the ground state from $\tilde{\theta}\in [\theta_c;\pi]$. Here, $\theta_c$ designates the location of the crossing point between this band and the line defining the chemical potential at $\mu=0$.
 Introducing the quantum Hall response of this eigenstate on the whole Brillouin zone with the topological invariant $\tilde{C}=1$ through the action of the time-reversal symmetry discussed earlier, we can verify that the missing term for
 $\tilde{\theta}\in [0;\theta_c]$ is precisely equal to the participation of the intermediate band polarized along $-z$ direction for $\tilde{\theta}\in [\pi-\theta_c;\theta_c]$
 \begin{equation}
\sigma_{xy}^{|-z\rangle,(2)} = \frac{e^2}{h}\int_0^{{\theta}_c} \sin\tilde{\theta} d\tilde{\theta} = \frac{e^2}{2h}(1-\cos{\theta}_c).
\end{equation}
The total quantum Hall response for the ground state is zero as in the quantum spin Hall effect and we can then introduce the {\it quantized $\mathbb{Z}_2$ topological invariant} for this topological semimetal (Fermi liquid) revealing the symmetry between electron and hole pockets 
\begin{equation}
\label{formula1}
\tilde{C}-C_{|-z\rangle} = \pm 2,
\end{equation}
for all parameters of $r$ within the domain $d_z-M<r<d_z+M$. We can also sum the quantum Hall responses introducing the {\it $\mathbb{Z}_2$ topological marker being non-quantized}
\begin{eqnarray}
\label{formula2}
\sigma_{xy}^{|-z\rangle,(1)}+ \sigma_{xy}^{|-z\rangle,(2)} = -\frac{e^2}{2h}(1+\cos \theta_c) = -\sigma_{xy}^{|+z\rangle}.
\end{eqnarray}
The angle $\theta_c$ varies continuously with the parameter $r$. Then, we can verify that $\theta_c=0$ when the intermediate band $|\psi_+\rangle\otimes|+z\rangle$
is fully occupied i.e. when $r<d_z-M$, reproducing the quantum spin Hall insulator. In that case, Eqs. (\ref{formula1}) and (\ref{formula2}) are equivalent since $\sigma_{xy}^{|-z\rangle,(2)}=0$.
Then, $\theta_c=\pi$ when the ground state becomes completely magnetized along $-z$ direction with $\sigma_{xy}^{|+z\rangle}=0$ i.e. when $r>d_z+M$. 
The two lowest-energy bands are then characterized with $\sigma_{xy}^{|-z\rangle,(1)}=-\sigma_{xy}^{|-z\rangle,(2)}$, which then reveals the quantum anomalous Hall $\mathbb{Z}$ structure of the system. 

In Fig. 1 (left), suppose we take a sample and decrease smoothly the chemical potential such that $\sigma_{xy}^{|-z\rangle,(2)}=0$ whereas $\sigma_{xy}^{|-z\rangle,(1)}=-\frac{e^2}{h}$. If we measure the conductance at the edges this is quantized (in units of $\frac{e^2}{h}$) for the spin-polarized fermion along $-z$ direction revealing the zero-energy {\it red mode} when adjusting the Fermi energy.  We can now increase smoothly the chemical potential such that the eigenstate $|\psi_+\rangle\otimes|+z\rangle$ is fully occupied in k-space. This gives rise to $\sigma_{xy}^{|+z\rangle}=\frac{e^2}{2h}\tilde{C}$ such that the {\it blue edge mode} structure also reveals $\tilde{C}=C_{|+z\rangle}=+1$ which is equal to $-C_{|-z\rangle}$ in the previous protocol. Below, Eqs. (\ref{formula1}) and (\ref{formula2}) will be revealed from circularly polarized light at half-filling, i.e. when $\mu=0$.

{\color{blue} Topological classification through circularly polarized light.---} 
Light-induced transitions resolved at specific points within the Brillouin zone are particularly meaningful, e.g. when the resonance is reached, to resolve topologically properties linked to the QSH effect \cite{Herrero,Light2}. These transitions select the same spin quantum
number. We reveal topological properties at half-filling from light. The light-matter coupling can be included in ${\cal H}({\bf k})$ as $\delta {\cal H}_{\pm}=A_0 e^{\pm i\omega t}\sigma^+\otimes\mathbb{I}+h.c.$ acting on the sublattices $|a\rangle=|+\rangle$, $|b\rangle=|-\rangle$. The resonance situation with each light polarization can be revealed through the transformation $|b\rangle \rightarrow e^{\mp \frac{i\omega t}{2}}|b'\rangle$ and $|a\rangle \rightarrow e^{\pm \frac{i\omega t}{2}}|a'\rangle$ such that $E_b-E_a=E_{|-\rangle}-E_{|+\rangle}=\pm \hbar\omega$; $\delta {\cal H}_{\pm}$ refers to the right-handed and left-handed light circular polarizations. The resonance situations are characterized through the inter-band transition probabilities $\Gamma_{\pm}(\omega)$ at $K$ and $K'$ 
\begin{equation}
\label{Gamma}
\Gamma_{\pm}(\omega) = \frac{2\pi}{\hbar}|\langle -| \delta {\cal H}_{\pm} | +\rangle|^2 \delta(E_{|-\rangle}-E_{|+\rangle}\mp \hbar\omega).
\end{equation}
If we select the $(-)$ yellow light polarization, fixing the resonance frequency such that $\hbar\omega=2|{\bf d}_+|=2(d_z+M)$,
this will mediate inter-band transitions at both Dirac points. Generalizing the approach of Ref. \cite{Light2}, this results in the following variation of ground-state population in time associated to $+z$ and $-z$ fermions
\begin{equation}
\label{probability}
\frac{d N_{|+z\rangle}}{dt} + \frac{d N_{|-z\rangle}}{dt} = -\pi \Delta \Gamma_- 
\end{equation}
where
\begin{eqnarray}
\Delta\Gamma_- &=& \frac{1}{2\pi}(\Gamma_-^{|+z\rangle}(K')+\Gamma_-^{|-z\rangle}(K)) \nonumber \\
&=& \frac{A_0^2}{\hbar^2}(\tilde{C}^2+C^2_{|-z\rangle}) =  \frac{A_0^2}{\hbar^2}(\tilde{C}-C_{|-z\rangle}).
\end{eqnarray}
At this stage, we introduce the frequency-integrated transition probabilities at $K$ and $K'$. Measuring the responses for each spin sector also allows us to verify the topological invariant e.g. $C_{|-z\rangle}$.
This responses agree with the QSH effect \cite{Light2} and information on the signs can be obtained from the direction of the photo-induced currents; if we do not integrate on light frequencies while remaining at resonance with the $K$ Dirac point, then the dynamics of ground-state population evolves as $t^2$.

Then, the geometrical approach on the sphere allowing us to reveal this information, which is developed for completeness in the Supplementary Material \cite{SM}, luckily also reveals the physics related to Eq. (\ref{formula2}). For this purpose, suppose we shine light on the sphere at the point with angle 
$\theta_c^+$ characterizing the crossing of $|\psi_-\rangle\otimes|+z\rangle$ with the Fermi energy. As shown in Fig. \ref{Figuretopo} (left), we can yet mediate inter-band transitions from this specific location in the Brillouin zone for the $+z$ spin polarization with the right-handed $(+)$ wave corresponding to the resonance frequency such that $\hbar\omega=2|{\bf d}_-|=2r$. The form factor entering into the specific inter-band transition probability $\langle \psi_-|\sigma_x|\psi_+\rangle \langle \psi_+|\sigma_x|\psi_-\rangle + \langle \psi_-|\sigma_y|\psi_+\rangle \langle \psi_+|\sigma_y|\psi_-\rangle$ at angle $\tilde{\theta}=\theta_c$ in Eq. (\ref{Gamma})
is precisely equal to the sum of the square of the conductivities 
$(\sigma_{xy}^{|+z\rangle})^2 + (\sigma_{xy}^{|-z\rangle,(2)})^2$ in Eq. (\ref{equationtransport}) forming the Fermi surface (in units of $\frac{h}{e^2}$). This is precisely related to the geometrical function $\alpha(\theta)$ \cite{Light2,Review}. This is similar as if a Fabry-Perot resonance would take place with light in momentum space \cite{SM}. With the same resonance frequency, the right-handed light wave can also promote transitions at $\pi-\theta_c$ i.e. around $K'$ in the Brillouin zone, from $|\psi_-\rangle\otimes|-z\rangle$ $|\psi_+\rangle\otimes|-z\rangle$, revealing the same form factor and the $\mathbb{Z}_2$ structure of Eq. (\ref{formula2}). 

{\color{blue} Zeeman effect and interactions.---} We can include interaction effects through the variational stochastic approach that we recently introduced for the interacting Haldane \cite{Light1} and Kane-Mele model \cite{HKH} where the energetics minimization principle then leads to the introduction of uniform (stochastic) variables $\phi_r = -\frac{1}{2}\langle S_r\rangle$ in the spin sector, where ${\bf S}=c^{\dagger}{\bf s} c$. For the Kane-Mele model, this analytical approach \cite{HKH} quantitatively agrees for the Mott transition line with Cluster dynamical mean-field theory \cite{WeiKarynStephan} and quantum Monte Carlo \cite{Wurzburg} already at a mean-field level with the choice of stochastic variables. At weak interactions, $\phi_x=\phi_y=0$ and the presence of Zeeman effects can lead to a finite value for $\phi_z$ such that this modify the $r$ term as $r\mathbb{I}\otimes s_z+U \phi_z \mathbb{I}\otimes s_z$. Then, $\phi_z$ can be evaluated from the band structure and net spin magnetization of the ground state such that $\phi_z=\frac{\pi}{v_F^2 N}(r^2-(d_z-M)^2)$ with $N$ being the number of sites. This is stable towards disorder which corresponds here to modify smoothly the term $M$ for each sample.

%\begin{figure*}[ht]\centering
%\includegraphics[width=1\textwidth]{model2withlight} 
%\vskip -0.3cm
%\caption{}
%\label{spectrawithlight}
%\end{figure*}

{\color{blue} Model with spin-dependent staggering potential.---} Here, we propose an alternative version of the model where the charge density wave substrate becomes an additional spin density wave layer e.g. an antiferromagnet on the honeycomb lattice, a N\' eel phase with local moments polarized along $z$ direction perpendicular to the plane. In that case, the itinerant electrons producing the Kane-Mele model will couple to these localized electrons through an Anderson model inducing then an antiferromagnetic Kondo coupling on each site. Recently, the occurrence of topological materials as a result of Kondo physics is attracting attention in the community with applications in heavy fermions and semimetals \cite{DzeroColeman,SmB6,KondoWeyl,Fabrizio}. In the present situation, since the local moments have a fixed net antiferromagnetic polarization along $z$ axis then the residual coupling on each site, i.e. the $M$ term becomes a spin-dependent staggering potential $\sum_{\bf k}\psi^{\dagger}({\bf k})(-M\sigma_z\otimes s_z)\psi({\bf k})$; we assume e.g. $M>0$. Then, we include uniform Zeeman effects through the $r$ term in the Hamiltonian. 
The interplay between the $M$ term and Zeeman effects yet allows for a topological semimetal when $d_z-M<r<d_z+M$; See Fig. 1, right.

The structure of the four eigenstates is similar to that for the model in Eq. (\ref{Hmodel}); see Table 1 second row. The main difference being that at the $K$ Dirac point, the energies just depend on $|{\bf d}_+|$ producing an inversion
between the two upper bands. In addition, within the same choice of parameters, the two lowest bands are now occupied similar to the QSH insulator. At the $K'$ Dirac point, the energies now depend
on $|{\bf d}_-|$ such that the second and third energy bands cross simultaneously the Fermi energy forming a semimetal ring when $|{\bf d_-}|=r$. This semimetallic Fermi liquid shows analogy with the QAH
semimetal in the band structure with the nodal ring \cite{HH,LeHurSaati1}, but the topological characterization of bands is distinct. The intermediate bands crossing the Fermi energy participate to the ground state quantum Hall responses 
in the same way as in Eqs. (\ref{formula1}) and (\ref{formula2}) \cite{SM}.

If we zoom on the band structure, there is a stable cross forming zero-energy modes which is another way to understand the $\mathbb{Z}_2$ character of this semimetal. Here, we relate this $\mathbb{Z}_2$ cross structure with the light responses at the $K$ Dirac point. To do so, we find it useful to build a correspondence with a model of two spheres or two interacting spins on the Bloch sphere that we recently introduced allowing for the definition of fractional topological numbers associated to half Skyrmions \cite{HH,KLH}. For the situation with a `mass' term $-M\sigma_z\otimes s_z$, the light response resolved in the two Dirac valleys yet reveals a  $\mathbb{Z}_2$ structure, and only one Dirac valley centered around $K$ responds to the light sources due to the fact that the light-matter coupling only involves pairs of bands with the same spin quantum number and describes transitions from occupied to empty bands at zero temperature. At the $K$ Dirac point, we obtain a similar equation as Eq. (\ref{probability}) with on the right-hand side a response function involving the two light polarizations
\begin{equation}
\Delta\Gamma =  \frac{1}{2\pi}(\Gamma_-^{|-z\rangle}(K)+\Gamma_+^{|+z\rangle}(K)) = \frac{A_0^2}{\hbar^2}(\tilde{C}-C_{|-z\rangle}).
\end{equation}
At the $K'$ Dirac point, $\Delta\Gamma=0$. If we average the responses at $K$ and $K'$ this reveals the halved $\mathbb{Z}_2$ number 
\begin{equation}
\frac{1}{2}(\tilde{C}-C_{|-z\rangle})) = \pm 1.
\end{equation}
The photo-induced currents then take an helical rotating structure similar to the QSH effect for this protocol in a topologically protected quantized way \cite{Light2,Review}.
The two filled or occupied bands are from one Dirac point equivalent to one pair of $\pm \pi$ Berry phases or one pair of $\pm \pi$ winding numbers corresponding to two half-Skyrmions with radial magnetic fields going in for one of this hemisphere and going out for the other; see Supplementary Material \cite{SM}. The light response then measures the same local marker characterizing the model of two spheres with the Hamiltonian $H_-$ \cite{HH}. The light response is also distinct for the QAH semimetal corresponding to a pair of $\pi$ Berry phases at the $K$ Dirac point or a pair of $\pi$ winding numbers \cite{LeHurSaati2}. 

{\color{blue} Conclusion.---} We have introduced topological semimetals in 2D i.e. topological Fermi liquids characterized through a $\mathbb{Z}_2$ invariant through microscopic models of band structures which can be realized, building a bridge between the quantum anomalous Hall insulator
and the quantum spin Hall insulator. The topological information can be resolved  and classified from circularly polarized light spectroscopy. The photo-induced responses encode signatures of protected helical electron channels and also of the Fermi surface.
\\
\\
{\it This work has benefitted from discussions with Sariah Al Saati and H\' el\` ene Bouchiat. This work is supported by the Deutsche Forschungsgemeinschaft (DFG), German
Research Foundation under Project No. 277974659.}

%\title{Topological Quantum Spin Hall Semimetals with Light}
%\author{Karyn Le Hur}
%\affiliation{CPHT, CNRS, École polytechnique, Institut Polytechnique de Paris, 91120 Palaiseau, France}

\onecolumngrid
\begin{center}
{\it Here, we derive a complete understanding of the topological band structures for the two models of crystals and give additional information on the responses to circularly polarized light related to geometrical functions}.
\end{center}
\maketitle

\onecolumngrid

\section{Topological Band Structure}

{\color{blue} Geometrical Representation on the Sphere.---} Related to Eqs. (1) and (2) in the Letter, it is useful to map the two-dimensional Brillouin zone on the surface of the sphere. To describe the eigenstates, we can also introduce a {\it spin-dependent} ${\bf d}$ vector:
\begin{equation}
\label{dvector}
(d_x,d_y,-\zeta d_z s_z+M) = |{\bf d}|\left(\sin\theta\cos\varphi,\sin\theta\sin\varphi,\cos\theta+\frac{M}{|{\bf d}|}\right)
\end{equation}
such that the Hamiltonian $H=\sum_{\bf k} \psi^{\dagger}({\bf k}){\cal H}({\bf k}) \psi({\bf k})$ in the pseudo-spin sector can be viewed as ${\bf d}(s_z)\cdot \mathbfit{\sigma}$. The Zeeman effects can be added as shifts $r s_z$ to the eigen-energies taking into account the spin structures of the eigenstates.
The angle $\theta\in[0;\pi]$ is the {\it polar} angle and $\varphi\in [0;2\pi]$ is the {\it equatorial} angle. The angle $\varphi$ on the sphere is then related to the polar angle $\tilde{\varphi}$ describing the dynamics around each Dirac point in the Brillouin zone.
Close to the Dirac points, the kinetic term components $d_x$ and $d_y$ take the forms  {\color{magenta} [1]}
\begin{eqnarray}
d_x &=& \hbar|{\bf p}| \cos\tilde{\varphi} \\ \nonumber
d_y &=& \hbar|{\bf p}| \sin(\zeta\tilde{\varphi}),
\end{eqnarray}
with $\zeta=+$ at the point $K$ and $\zeta=-1$ at the point $K'$; ${\bf p}$ corresponds to a small {\it wave-vector} deviation from each Dirac point such that $d_x=d_y=0$ at $K$ and $K'$. The two inequivalent Dirac points within the Brillouin zone correspond to the vectors ${\bf K}=(\frac{2\pi}{3a},\frac{2\pi}{3\sqrt{3}a})$ and ${\bf K'}=(\frac{2\pi}{3a},-\frac{2\pi}{3\sqrt{3}a})$; here, $a$ is the lattice spacing. 
To study the light-matter interaction, it is particularly useful to develop the responses close to the Dirac points. The $K$ Dirac point is e.g. at the {\it north} pole and $K'$ at {\it south} pole and we can then formulate a simple map such that 
$k_y=\frac{2\pi}{3\sqrt{3}a} -\frac{4}{3\sqrt{3}a}\theta$ {\color{magenta} [1]}. 
This also allows for a correspondence on the cylinder.

We can absorb the (mass) term $M$ into a redefinition of the polar angle 
\begin{equation}
\tan\tilde{\theta}=\frac{\sin\theta}{\cos\theta+\frac{M}{|{\bf d}|}}.
\end{equation}
When the parameter $M$ satisfies $-|{\bf d}|<M<|{\bf d}|$ then $\tilde{\theta}\in [0;\pi]$. 
Therefore, this leads to the correspondence
\begin{equation}
\label{number}
\tan\tilde{\theta} = \frac{\hbar v_F |{\bf p}|}{-d_z s_z \zeta+M}.
\end{equation}

{\color{blue} Characterization of Topological Band Structure.---} To satisfy that $\tilde{\theta}\in [0;\pi]$, if we are in the $K$ valley, this implies that the lowest energy band is spin-polarized along $s_z=-1$. In this case, associated to the energy $-|{\bf d}_+|=-\sqrt{d_x^2 + d_y^2 + (d_z+M)^2}$
in the pseudo-spin sector the eigenstate is $|\psi_-\rangle = -\sin\frac{\tilde{\theta}}{2}e^{-i\frac{{\varphi}}{2}}|+\rangle + \cos\frac{\tilde{\theta}}{2}e^{i\frac{{\varphi}}{2}}|-\rangle$. Including Zeeman effect, 
the eigenstate {\color{blue} $|\psi_-\rangle\otimes|-z\rangle$} has energy $-|{\bf d}_+|-r$ in the $K$ valley, equivalent to the {\it lowest-energy} band, and becomes the {\it second-energy} band in the $K'$ valley with energy $-|{\bf d}_-|-r$ such that ${\bf d}_-=\sqrt{d_x^2+d_y^2+(d_z-M)^2}$.
Going from one valley to another is equivalent to flip the sign of $\zeta$, and to modify the direction of the azimuthal angle which does not modify physical properties. This is also equivalent indeed to modify $\tilde{\theta}\rightarrow \pi$. Since we do not modify the sign of $M$ in this transformation $K\rightarrow K'$, this implies that the eigenstates energies are modified according to $|{\bf d}_{\pm}|\rightarrow |{\bf d}_{\mp}|$. The pseudo-spin polarization turns from $|-\rangle$ to $|+\rangle$ when going from $K$ to $K'$. 
We can introduce a topological number similarly as in cartesian coordinates for this eigenstate associated to the ground-state defined in terms of the  Berry curvature ${F}_{\tilde{\theta}\varphi}=\partial_{\tilde{\theta}}{\cal A}_{\varphi}=-\frac{\sin\tilde{\theta}}{2}$ {\color{magenta} [2,3]}
\begin{equation}
C_{|-z\rangle} = \frac{1}{2\pi}\int_0^{\pi} \int_0^{2\pi} {F}_{\tilde{\theta}\varphi}d\tilde{\theta}d\varphi = {\cal A}_{\varphi}(\pi)-{\cal A}_{\varphi}(0)=-1.
\end{equation}
We introduce the Berry gauge potential ${\cal A}_{\varphi}=-i\langle \psi| \partial_{\varphi} |\psi\rangle$ which is equal to $\frac{\cos\tilde{\theta}}{2}$ when $|\psi\rangle = |\psi_-\rangle$.
This results in the quantum Hall conductivity ${\color{blue} \sigma_{xy}^{|-z\rangle}=\frac{e^2}{h}C_{|-z\rangle}}$ which agrees with a direct evaluation from the Kubo formalism {\color{magenta} [3,4,5]}. 

The {\it second-energy} band at the $K$ Dirac point is related to the first one through the transformation $s_z\rightarrow -s_z$. To satisfy that Eq. (\ref{number}) remains applicable such that $\tilde{\theta}\in [0;\pi]$ this also implies to flip the role of ${\bf d}\rightarrow -{\bf d}$
which is also equivalent to modify $\sigma_z\rightarrow -\sigma_z$ around the two Dirac points. The eigenstate in the pseudo-spin sector then turns from $|\psi_-\rangle\rightarrow |\psi_+\rangle = \cos\frac{\tilde{\theta}}{2}e^{-i\frac{{\varphi}}{2}}|+\rangle + 
\sin\frac{\tilde{\theta}}{2}e^{i\frac{{\varphi}}{2}}|-\rangle$. This reveals the same $\mathbb{Z}_2$ symmetry between the two lowest-energy bands as in the Kane-Mele model hidden in the transformation, ${\color{blue} s_z\rightarrow -s_z}$
and ${\color{blue}\sigma_z\rightarrow -\sigma_z}$. The modification $s_z\rightarrow -s_z$ then implies that the corresponding energy at the $K$ point for this eigenstate ${\color{blue} |\psi_+\rangle\otimes|+z\rangle}$ is $-|{\bf d}_-|+r$ whereas at the $K'$ point it becomes $-|{\bf d}_+|+r$ corresponding to the {\it lowest-energy} band. The pseudo-spin polarization then turns  from $|+\rangle$ to $|-\rangle$ when going from $K$ to $K'$. Since the two lowest-energy bands have orthogonal spin eigenstates it is possible to integrate the Berry curvature
associated to each band in a domain associated to the ground state at zero temperature. If we fix within the band theory, the chemical potential  at $\mu=0$ corresponding to half-filling with the range of parameters ${\color{blue} d_z-M<r<d_z+M}$, then the energy band associated to
$|\psi_+\rangle\otimes|+z\rangle$ is above the Fermi energy for $\tilde{\theta}\in [0;\theta_c[$ and below the Fermi energy for $\tilde{\theta}\in ]\theta_c;\pi]$. We remind that $d_x,d_y\rightarrow 0$ when $\theta\rightarrow 0,\pi$.
The Berry curvature $F_{\tilde{\theta}\varphi}$ associated to $|\psi_+\rangle$ is $\frac{\sin\tilde{\theta}}{2}$ such that this results in the conductivity response for the ground state 
\begin{equation}
{\color{blue}
\sigma_{xy}^{|+z\rangle} = \frac{e^2}{2h}\int_{\theta_c}^{\pi} \sin\tilde{\theta}d\tilde{\theta} = \frac{e^2}{2h}(1+\cos\theta_c).
}
\end{equation}

At the $K'$ Dirac point, there is also an electron pocket crossing the Fermi energy associated to the eigenstate ${\color{blue} |\psi_+\rangle\otimes|-z\rangle}$ in the domain of angle $\tilde{\theta}\in [\pi-\theta_c;\pi]$. It is interesting to observe that this eigenstate or electron pocket can be precisely teleported around the $K$ point through time-reversal symmetry corresponding to rotate the spin eigenstate (with a rotation axis corresponding to $y$ direction) from $|-z\rangle$ to $|+z\rangle$. Time-reversal symmetry is also equivalent to modify the sign of the momentum or wave-vector corresponding here to swap the role of $K'$ and $K$. 
Under time-reversal symmetry, this is then equivalent to an eigenstate $|\psi_+\rangle\otimes|+z\rangle$ at the $K$ point but with a corresponding energy $|{\bf d}_-|-r$ i.e. below the Fermi energy. Saying this is equivalent to justify that adding the portion of Fermi surface close to $K'$ participating in the ground-state is equivalent to {\color{blue} a Berry curvature linked to $|\psi_+\rangle\otimes|+z\rangle$ integrated on $\tilde{\theta}\in [0;\pi]$ which should then reveal a (protected) quantized quantum Hall conductivity for the Fermi surface}. This can be shown as follows. The corresponding Fermi surface Hall conductivity in momentum space close to $K'$ 
from the red dashed metallic region
can be rephrased as
{\color{blue}
\begin{equation}
\sigma_{xy}^{|-z\rangle,(2)} = \frac{e^2}{2h}\int_{\pi-\theta_c}^{\pi} \sin\tilde{\theta}d\tilde{\theta} = \frac{e^2}{2h}\int_{0}^{\theta_c} \sin\tilde{\theta}d\tilde{\theta} = \sigma_{xy}^{|+z\rangle,(2)} = \frac{e^2}{2h}(1-\cos\theta_c).
\end{equation}
}
This is then equivalently re-written as the quantum Hall response of the teleported orange dotted Fermi arc close to the $K$ Dirac point.
We can then verify the quantization coming from the Fermi surface giving rise to a $\mathbb{Z}$ invariant:
{\color{blue}
\begin{equation}
\sigma_{xy}^{|+z\rangle} +  \sigma_{xy}^{|+z\rangle,(2)} = \frac{e^2}{h}\tilde{C}.
\end{equation}
}
Since the lowest-energy bands $|\psi_+\rangle\otimes|+z\rangle$ and $|\psi_-\rangle\otimes|-z\rangle$ are precisely related through the $\mathbb{Z}_2$ symmetry $\sigma_z\rightarrow -\sigma_z$ and $s_z\rightarrow -s_z$, this justifies the
characterization of the band structure through a quantized $\mathbb{Z}_2$ topological invariant
{\color{blue}
\begin{equation}
\label{number2}
\tilde{C}-C_{|-z\rangle} = \pm 2.
\end{equation}
}
We also verify that $\sigma_{xy}^{|-z\rangle}+\sigma_{xy}^{|-z\rangle,(2)} = - \sigma_{xy}^{|+z\rangle} =  -\frac{e^2}{2h}(1+\cos\theta_c)$. The total quantum Hall conductivity is zero for the ground-state. Eq. (\ref{number2}) justifies the introduction of the term quantum spin Hall semimetal or quantum spin Hall 
$\mathbb{Z}_2$ Fermi liquid to characterize or designate this system. This ground-state has also a net magnetization (related to the variable $\phi_z$ in the Letter, page 4) therefore it is a topological quantum spin Hall magnetic semimetal. The presence of the Zeeman effect has consequences e.g. it implies that the energy spectrum is not invariant under the reversal of ${\bf k}\rightarrow -{\bf k}$, which has allowed us to teleport an additional Fermi region at the $K$ point below the Fermi energy.

It is relevant to mention that the quantization of the Fermi surface for this topological semimetal (Fermi liquid) is distinct compared to the classification of the quantum anomalous Hall topological semimetal on the honeycomb lattice {\color{magenta} [2-5]}. 

%The magnetic field is applied along the direction of the plane e.g. along $x$ direction. In that case, the eigenstates forming the intermediate bands $2$ and $3$ reveal the structure of eigenstates: at K) $|\psi_2\rangle = |\psi_-\rangle\otimes|+_x\rangle$, $|\psi_3\rangle = |\psi_+\rangle\otimes|-_x\rangle$ % and at K') $|\psi_2\rangle = |\psi_+\rangle\otimes|-_x\rangle$, $|\psi_3\rangle = |\psi_-\rangle\otimes|+_x\rangle$. Therefore, if we take the band $2$ just below the Fermi energy, the regions close to $K$ and $K'$ are now related through a $\mathbb{Z}_2$ structure i.e. when modifying simultaneously % $s_x\rightarrow -s_x$ and $\sigma_z\rightarrow -\sigma_z$ which justifies the introduction of a $\mathbb{Z}_2$ topological invariant characterizing the Fermi surface formulated as $C_{-,|+x\rangle}-C_{+,|-x\rangle}$. This transformation is also related to the Pfaffian in the Kane-Mele model because % rotating $s_x\rightarrow -s_x$ is also equivalent to act with the time-reversal symmetry {\color{magenta} [6]}. The lowest-energy band is characterized through a $\mathbb{Z}$ topological invariant $C_{-,|-x\rangle}$ similar to the Haldane model such that {\color{magenta} [4]}
% {\color{blue} 
% \begin{equation}
% C_{-,|+x\rangle}-C_{+,|-x\rangle}=C_{-,|-x\rangle}.
% \end{equation}
%}

\section{Light-Matter Interaction}

{\color{blue} Light-Matter Interaction and Inter-Band Transition Probabilities.---} Including the light-matter interaction, this modifies the kinetic terms $d_x$ and $d_y$ in the Hamiltonian, equally for the two spin polarizations such that {\color{magenta} [1]}
\begin{equation}
\delta {\cal H}_{\pm} = (A_0 e^{\pm i \omega t} \sigma^+ +h.c.)\otimes \mathbb{I}.
\end{equation}
%Within the Dirac approximation, the light-matter interaction gives rise to a vector potential which  is equivalent to a boost of momentum. 
The resonance situation is obtained when modifying the pseudo-spin eigenstates $|+\rangle\rightarrow e^{\pm i \frac{\omega t}{2}} |+'\rangle$ and $|-\rangle\rightarrow e^{\mp i \frac{\omega t}{2}} |-'\rangle$ such that $(E_- - E_+)=\pm \hbar\omega$.
This gives rise to the inter-band  transition probabilities
\begin{equation}
\label{probabilities}
\Gamma_{\pm}(\omega) = \frac{2\pi}{\hbar}|\langle \psi_- | \delta {\cal H}_{\pm} |\psi_+\rangle|^2 \delta(E_- - E_+ \mp \hbar\omega).
\end{equation}
These light-induced optical transitions imply that the same {\it spin-polarization} is preserved within the protocole and imply flipping the pseudo-spin. The chirality of the wave, right or left, is simply fixed with the energy conservation e.g. a transition from sublattice $B$ to sublattice $A$ $|-\rangle\rightarrow |+\rangle$ is realized through a left-moving circularly polarized wave. Similarly, a right-moving wave allows to perform transitions from sublattice $A$ to sublattice $B$
$|+\rangle\rightarrow |-\rangle$. It is also important to observe that $|\psi_+(K)\rangle=|+\rangle$ and $|\psi_-(K)\rangle=|-\rangle$ with $|\psi_+(K)\rangle=|\psi_-(K')\rangle$ and $|\psi_+(K')\rangle=|\psi_-(K)\rangle$. Eq. (\ref{probabilities}) 
measures the inter-band transition probabilities which are directly related to the variations in time of the ground-state populations associated to a spin polarization. These inter-band transition probabilities are also directly related to the photo-induced currents through conservation laws {\color{magenta}[1]}. 

{\color{blue} Geometrical Responses Encoded in Light.---} Through the formalism on the sphere, it is then possible to derive interesting general relations such as
\begin{equation}
\langle \psi_-| \sigma_x | \psi_+\rangle \langle \psi_+| \sigma_x |\psi_-\rangle = \sin^4\frac{\tilde{\theta}}{2} +  \cos^4\frac{\tilde{\theta}}{2} - 2 \sin^2\frac{\tilde{\theta}}{2} \cos^2\frac{\tilde{\theta}}{2} \cos\varphi.
\end{equation}
Similarly, we obtain
\begin{equation}
\langle \psi_-| \sigma_y | \psi_+\rangle \langle \psi_+| \sigma_y |\psi_-\rangle = \sin^4\frac{\tilde{\theta}}{2} +  \cos^4\frac{\tilde{\theta}}{2} + 2 \sin^2\frac{\tilde{\theta}}{2} \cos^2\frac{\tilde{\theta}}{2} \cos\varphi.
\end{equation}
Therefore, the light-matter interaction gives rise to a response function which is directly proportional to {\color{magenta} [1,3]}
\begin{equation}
{\color{blue}
\alpha(\tilde{\theta}) = \left(\cos^4\frac{\tilde{\theta}}{2} + \sin^4\frac{\tilde{\theta}}{2}\right).
}
\end{equation}
At this stage, it is not so immediate to see how this function reveals a topological information in terms of the Berry functions ${\cal A}_{\varphi}$ and $F_{\tilde{\theta}\varphi}$; it is however already clear that this function is equal to unity at the poles or at the Dirac points. It is then useful to encode this function into {\it new geometrical functions}, as follows. 

When the chemical potential is zero, at half-filling, the left-handed polarized wave $(-)$ gives rise to transitions between ${\color{blue} |\psi_-(K)\rangle\otimes|-z\rangle \rightarrow |\psi_+(K)\rangle\otimes|-z\rangle}$ at the $K$ point and between ${\color{blue} |\psi_+(K')\rangle\otimes|+z\rangle \rightarrow |\psi_-(K')\rangle\otimes|+z\rangle}$ at the $K'$ point. The lowest energy band at the $K$ Dirac point is characterized through the Berry gauge field ${\cal A}_{\varphi}(\tilde{\theta})=\frac{\cos\tilde{\theta}}{2}$. Suppose we measure the light responses at the angle $\theta^*$. Then, it is particularly relevant to introduce the functions
{\color{magenta} [1-3]}
\begin{eqnarray}
{\cal A}'_{\varphi}(\tilde{\theta}<\theta^*) &=& {\cal A}'_{\varphi}(\tilde{\theta}=\theta^* - \epsilon)={\cal A}'_{\varphi}(\tilde{\theta}=\theta^*-\epsilon)-{\cal A}_{\varphi}(0) = -\sin^2\frac{\theta^*}{2} \nonumber \\
{\cal A}'_{\varphi}(\tilde{\theta}>\theta^*) &=& {\cal A}'_{\varphi}(\tilde{\theta}=\theta^* + \epsilon)={\cal A}'_{\varphi}(\tilde{\theta}=\theta^*+\epsilon)-{\cal A}_{\varphi}(\pi) = \cos^2\frac{\theta^*}{2}.
\end{eqnarray}
The field ${\cal A}'_{\varphi}$ is discontinuous at the interface i.e. here at the point along the path where we measure the light response, and we also have the important identities ${\cal A}'_{\varphi}(\tilde{\theta}<\theta^*)=0$ when $\theta^*=0$ and ${\cal A}'_{\varphi}(\tilde{\theta}>\theta^*)=0$ when $\theta^*=\pi$. We can then verify that ${\cal A}'_{\varphi}(\tilde{\theta}<\theta^*) - {\cal A}'_{\varphi}(\tilde{\theta}>\theta^*) = {\cal A}_{\varphi}(\pi)-{\cal A}_{\varphi}(0)=C_{|-z\rangle}$. In this way, for the spin polarization $|-z\rangle$, the function $\alpha_{|-z\rangle}$ then becomes equivalent to {\color{magenta} [1,3]}
\begin{equation}
{\color{blue}
\alpha_{|-z\rangle} = C_{|-z\rangle}^2 + 2{\cal A}'_{\varphi}(\tilde{\theta}<\theta^*){\cal A}'_{\varphi}(\tilde{\theta}>\theta^*).
}
\end{equation}
Therefore, when $\theta^*=0$ we have ${\color{blue} \alpha_{|-z\rangle} = C_{|-z\rangle}^2}$. This justifies why the left-handed wave probes the square of the topological number associated to the lowest
energy band locally at the $K$ Dirac point. This supposes that light is at resonance with the energy gap i.e. $\hbar\omega=2|{\bf d}_+|$. 

For the transition $|\psi_+(K')\rangle\otimes|+z\rangle \rightarrow 
|\psi_-(K')\rangle\otimes|+z\rangle$ involving the spin polarization $|+z\rangle$, we can then invoke the correspondence $|\psi_+(K')\rangle\rightarrow |\psi_-(K)\rangle$ and $|\psi_-(K')\rangle\rightarrow |\psi_+(K)\rangle$
such that the light-response for the $|+z\rangle$ polarization should be identical to the one for the $|-z\rangle$ polarization with the prerequisite that we can introduce the topological invariant $\tilde{C}$ defined on the whole interval 
$\tilde{\theta}\in[0;\pi]$ (introduced from time-reversal symmetry). Therefore, the same light polarization $(-)$ with the same resonance frequency corresponding to $\hbar\omega=2|{\bf d_+}|$ will probe the topological invariant ${\color{blue} \alpha_{|+z\rangle} = \tilde{C}^2}$. We have justified in the preceding Section that the introduction of $\tilde{C}$ is also equivalent to a $\mathbb{Z}_2$ symmetry $\sigma_z\rightarrow -\sigma_z$, $s_z\rightarrow -s_z$ similar to the
quantum spin Hall effect. If we would modify the sign of $d_z$, then this corresponds to modify $\sigma_z\rightarrow -\sigma_z$ which is equivalent to modify $|\psi_-\rangle \leftrightarrow |\psi_+\rangle$ and therefore to modify the
circulation of the electromagnetic wave. 

If we measure the responses to the right-handed light polarization at $\theta^*= \theta_c$, this reveals the same information as the quantum Hall conductivities from the partially filled regions since we can also identify
{\color{blue}
\begin{eqnarray}
\alpha_{|+z\rangle}(\theta_c) = \frac{1}{4}(1+\cos\theta_c)^2 + \frac{1}{4}(1-\cos\theta_c)^2 = \left(\sigma_{xy}^{|+z\rangle}\frac{h}{e^2}\right)^2 + \left(\sigma_{xy}^{|-z\rangle,(2)}\frac{h}{e^2}\right)^2.
\end{eqnarray}
}
This requires to adjust the resonance frequency such that $\hbar\omega=2|{\bf d}_-|=2r$.

\section{Spin-Dependent Potential and Topological Characterization through Light}

{\color{blue} Topological Characterization of Band Structure.---} For the situation with a spin-dependent potential which can be realized e.g. through a substrate with a spin density wave, the spin-dependent ${\bf d}$ vector turns into
\begin{equation}
(d_x,d_y,-\zeta d_z s_z-Ms_z) = |{\bf d}|\left(\sin\tilde{\theta}\cos\varphi,\sin\tilde{\theta}\sin\varphi,\cos\tilde{\theta}\right).
\end{equation}
The main difference is that at the $K$ Dirac point or at $\tilde{\theta}=0$ the energies depend on $|{\bf d_+}|$ such that the two lowest energy bands with the same quantum numbers as above remain below the Fermi energy.
The two upper bands are above the Fermi energy at half-filling. These pairs of bands are characterized through the $\mathbb{Z}_2$ symmetry $s_z\rightarrow -s_z$, $\sigma_z\rightarrow -\sigma_z$. Within this version of the model, the energies around $K'$ depend on $|{\bf d_-}|$ such that the eigenstate $|\psi_-\rangle\otimes|-z\rangle$ remains below the Fermi energy encoding the same topological invariant $C_{|-z\rangle}$. The eigenstate $|\psi_+\rangle\otimes|+z\rangle$ now crosses the Fermi energy giving rise to a ring structure whereas there is an additional Fermi surface forming the ground-state with spin polarization $|-z\rangle$ and corresponding eigenstate $|\psi_+\rangle\otimes|-z\rangle$ for $\tilde{\theta}\in[\theta_c';\pi]$ with $\theta_c'=\pi-\theta_c$. These two energy bands give rise to the quantum Hall responses
{\color{blue}
\begin{equation}
\sigma_{xy}^{|+z\rangle} = \frac{e^2}{h}\int_0^{\theta_c'} \frac{\sin\tilde{\theta}}{2}d\tilde{\theta} = -\frac{e^2}{2h}(\cos\theta_c'-1)
\end{equation}
\begin{equation}
\sigma_{xy}^{|-z\rangle,(2)} = \frac{e^2}{h}\int_{\theta_c'}^{\pi} \frac{\sin\tilde{\theta}}{2}d\tilde{\theta} = \frac{e^2}{2h}(1+\cos\theta_c'),
\end{equation}
}
such that they provide the same classification as the system above, Therefore, this results in the same {\it quantized $\mathbb{Z}$ topological invariant} of the Fermi surface which is formulated here as {\color{blue} $\sigma_{xy}^{|+z\rangle} + \sigma_{xy}^{|-z\rangle,(2)} = \frac{e^2}{h}\tilde{C}$ with 
$\tilde{C}=+1$.} Eqs. (6) and (7) in the Letter are yet applicable. 

{\color{blue} Halved $\mathbb{Z}_2$ Invariant and the Models with Two Spheres.---} The topological characterization through light can be equivalently introduced as a halved $\mathbb{Z}_2$ invariant locally in the Brillouin zone. Due to the fact that the eigenstate $|\psi_+\rangle\otimes|+z\rangle$ at $K'$ is above the Fermi energy, at half-filling, there is no possible light-induced inter-band transition in this valley since we are in the presence of two pairs of occupied or empty bands. On the other hand, around the $K$ Dirac point, we can reveal the $\mathbb{Z}_2$ topological invariant as follows. Similarly as the situation above, at the $K$ point the left-handed light $(-)$ polarization will reveal ${\color{blue} \alpha_{|-z\rangle}=C_{|-z\rangle}^2}$ whereas the right handed $(+)$ light polarization can now induce transitions between states $|\psi_+\rangle\otimes|+z\rangle\rightarrow |\psi_-\rangle\otimes|+z\rangle$ with the same frequency resonance $\hbar\omega=2|{\bf d}_+|$ measuring locally the topological marker ${\color{blue} \alpha_{|+z\rangle}=\tilde{C}^2}$. The structure of the photo-induced currents at the $K$ Dirac point is identical to those for the quantum spin Hall effect. If we average the light responses on the two Dirac valleys, then this is equivalent to introduce the topological halved marker
{\color{blue}
\begin{equation}
\label{numbercorrect}
\frac{1}{2}(\tilde{C}-C_{|-z\rangle}) = \pm 1.
\end{equation}
}
From the description of the topological invariant from the poles in terms of the Berry gauge potentials, this number is also equivalent to a pair of $\pm \pi$ Berry phases or winding numbers encoding the symmetry
$\sigma_z\rightarrow -\sigma_z$, $s_z\rightarrow -s_z$. This is also equivalent to a pair of half Skyrmions, one with the topological number $\frac{1}{2}$ and one with the topological number $-\frac{1}{2}$.
For a comparison, the quantum anomalous Hall semimetal is characterized through a pair of $\pi$ Berry phases {\color{magenta} [2,3,5]}.

We can also build a parallel of this invariant in terms of a model of two interacting spins on a sphere or on a cylinder. Suppose two particles $i=1,2$ interacting at the two poles through a model {\color{magenta} [2,5]}
\begin{equation}
H = -H\zeta \sigma_z^1 -M \sigma_z^1 +H \zeta \sigma_z^2 + M \sigma_z^2 - r \sigma_z^1 \sigma_z^2,
\end{equation}
where $\zeta=\pm 1$ at the two poles respectively. When $\zeta=+1$, at the north pole, the ground state is $|+\rangle\otimes|-\rangle$ with energy $-2(H+M)+r$. It is the ground state as long as $r<H+M$. When $\zeta=-1$,
the ferromagnetic interaction favors the two triplet states $|+\rangle\otimes |+\rangle$ and $|-\rangle\otimes|-\rangle$ with energy $-r$. This is energetically preferable as long as $r>H-M$.  Quantum mechanical effects from perturbation theory around south pole i.e. including small transverse field effects 
$\alpha_i \sigma_x^i$ with $\alpha_i$ infinitesimal close to the poles will entangle these two classical ground states through an operator 
$\sim - \frac{\alpha_1 \alpha_2}{r} \sigma_x^1 \sigma_x^2$. The ground state then becomes $|\psi\rangle = \frac{1}{\sqrt{2}}(|+ +\rangle + |- - \rangle)$. If we identify $1=|+z\rangle$ and $2=|-z\rangle$, then the two-particles ground state $|+\rangle\otimes|-\rangle$ is identical to the one in the lattice model in the $K$ valley. At south pole, we have $\langle \sigma_z^i(\pi)\rangle=0$. 
The topological number for each particle $i=1,2$ can be then formulated as {\color{magenta} [2,3,5]}
{\color{blue}
\begin{equation}
C_i = \frac{1}{2}(\langle \sigma_z^i(\tilde{\theta}=0)\rangle - \langle \sigma_z^i(\tilde{\theta}=\pi)\rangle) = \frac{1}{2}\langle\sigma_z^i (0)\rangle,
\end{equation}
}
implying that $C_1-C_2=\pm 1$ similarly to Eq. (\ref{numbercorrect}). An important symmetry for this model is the $\mathbb{Z}_2$ inversion $\sigma_z^i \rightarrow -\sigma_z^i$.
The classification with $C_1+C_2 = \pm 1$ on the sphere or cylinder is equivalent to the quantum anomalous semimetal on the lattice {\color{magenta} [2-5]}.

\end{document}